\documentclass[sigconf,nonacm]{acmart}

\usepackage{booktabs}
\usepackage{multirow}
\usepackage{xcolor}
\usepackage{tikz}
\usepackage{booktabs}
\usepackage{tabularx}
\usepackage{array}
\usetikzlibrary{arrows.meta,positioning,calc,fit}

\newcommand{\sys}{\textsc{AgentCount}}

\begin{document}

\title{\sys{}: An Agentic Workflow for Traceable Colony Counting and Insight Generation from Agar Plate Images}

\author{Zhuosheng Liu}
\authornote{Both authors contributed equally to this research.}
\authornote{Corresponding author.}
\affiliation{%
  \institution{University of California, Davis}
  \city{Davis}
  \state{CA}
  \country{USA}}
\email{zsliu@ucdavis.edu}
\author{Austin E. Weber}
\authornotemark[1]
\affiliation{%
  \institution{University of California, Davis}
  \city{Davis}
  \state{CA}
  \country{USA}}
\email{aeweber@ucdavis.edu}

\begin{abstract}
Colony counting is commonly used to evaluate microbial populations in food safety testing and microbiological studies. Despite its wide application, manual counting is time-consuming and operator-dependent, and the final count often lacks a record of the objects included in the measurement. To address these limitations, this study presents \sys{}, an open-source agentic workflow for colony counting and result interpretation from agar plate images. Plate photographs and plating metadata are processed through seven agents for image quality control, plate localisation, colony detection, counting, CFU calculation, validation, and reporting. Colony detection is performed using Segment Anything Model~3 (SAM\,3) with multiple text prompts and morphology-based filtering. After that, replicate statistics are calculated and a large language model (LLM)-based agent is used to generate a results narrative. Intermediate outputs, including quality scores, segmentation masks, detections, and warnings, are retained to support review of the reported values. The workflow also extends to skill-guided analysis of archived microbial population data. In a retrospective comparison with two previously published studies, we benchmarked 241 published endpoints, including all 216 mean$\pm$SD values of the meat-analogue pathogen tables; on the 168 endpoints with a complete $n{=}6$ replicate source, 164 (97.6\%) agreed within 0.01 log$_{10}$ units. Overall, \sys{} demonstrates the promising potential of agentic workflows for applications in applied microbiology.

\end{abstract}

\keywords{LLM agents, scientific workflows, foundation segmentation models, colony counting, data provenance}

\maketitle

%% =====================================================================
\section{Introduction}
%% =====================================================================
Plate counting is one of the most commonly used methods for evaluating microbial populations in food safety testing, clinical microbiology, antimicrobial screening, and environmental monitoring. Plating is among the most common methods for studying microbial fitness and survival~\cite{liu2024vpreview}, as in studies of foodborne pathogen survival on dried apricots made with and without sulfur dioxide treatment~\cite{liu2021apricots}, \textit{Listeria monocytogenes} survival on growing and harvested mushrooms~\cite{tu2025mushrooms}, microbial control during peach drying and storage~\cite{liu2022peaches}, bacterial growth in meat and plant-based alternatives~\cite{liu2023plantmeat}, and the fitness of \textit{Vibrio parahaemolyticus} in seawater at different shellfish harvesting temperatures~\cite{liu2023vprnaseq}. The labor capital requirement to support these investigations, proportional to the total number of inoculated plates, scales multiplicatively as a function of increasing numbers of experimental conditions, sampling time points, selective media, dilution levels, and biological or technical replicates. Manual plate counting is still widely used, but  comes with a substantial time and labor cost, with operator error introducing additional variation in the results. Additionally, this procedure frequently records only a final summated count, which provides limited traceability and detailing regarding which objects were counted, and why a plate was accepted or rejected for further analysis. 

\sys{}, an open-source agentic workflow, connects image analysis, CFU calculation, statistics, and reporting using existing plate photographs and compatible computing resources, potentially avoiding the purchase of a dedicated colony counter where suitable imaging and compute are already available. Net savings depend on acquisition, compute, and review costs; matched end-to-end timing and cost measurements remain necessary. \sys{} is open source and available on GitHub at \url{https://github.com/jlk666/AgentCount}.

Automation of colony counting using image processing and machine learning has been frequently explored in the field of microbiology. Classical tools, such as OpenCFU~\cite{geissmann2013opencfu} and NICE~\cite{clarke2010nice}, which employ intensity thresholding and watershed-based segmentation methods, respectively. These approaches perform well under controlled imaging conditions, while variation in lighting, medium colour, and colony morphology can affect their performance. Supervised detectors~\cite{ferrari2017bacterial} provide an alternative methodology, however annotated images are needed for model training and adaptation to new conditions. Further, image-based counting alone does not address the subsequent steps of quality assessment, CFU calculation, replicate analysis, and result interpretation.

With the advances in foundation segmentation models, promptable methods~\cite{kirillov2023sam} provide an opportunity to analyze plate images without task-specific annotation. SAM\,3~\cite{carion2025sam3} supports open-vocabulary concept prompts, such as ``small circular colonies on TSA agar'', for zero-shot segmentation. Agentic frameworks allow different analytical steps to be organized into a shared workflow with explicit inputs and outputs.

Therefore, the goal of this study is to develop a traceable workflow that connects plate image analysis with quantitative reporting. The specific components demonstrated in \sys{} are:

\begin{itemize}
  \item \textbf{An integrated agentic workflow} for quality control, plate localization, detection, counting, CFU calculation, validation, and reporting using a shared typed state.
  \item \textbf{Colony detection without task-specific annotation} using SAM\,3, multiple text prompts, contrast enhancement, and morphology-based filtering.
  \item \textbf{Traceable, automated analysis beyond counting}, linking every reported value to intermediate outputs and machine-readable warnings, and combining batch statistics, LLM-assisted interpretation, and skill-guided archive reanalysis evaluated against two previously published studies.
  \item \textbf{A public agar plate benchmark} of fifty-seven in-house images across MAC, MRS, PCA, and VRBA media with manual counts and countability labels, released to evaluate classification and counting accuracy.

\end{itemize}

%% =====================================================================
\section{System Overview}
%% =====================================================================
\begin{figure*}[t]
  \centering
  \includegraphics[width=\textwidth]{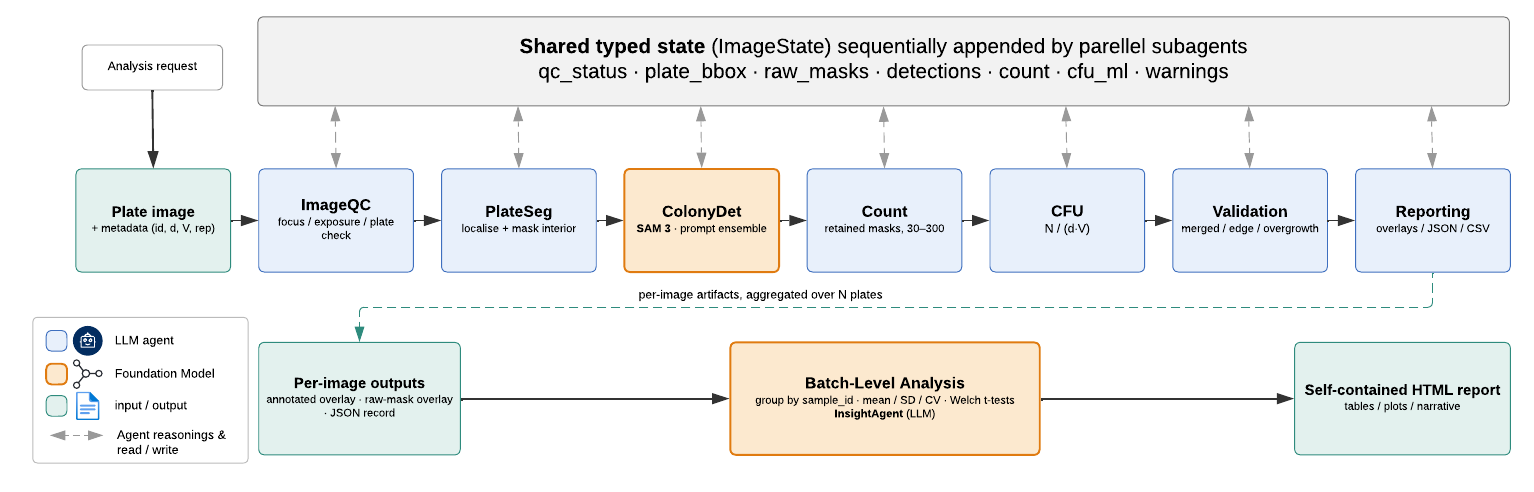}
  \caption{\sys{} architecture. Seven per-image agents (foundation-model agents in orange) are nodes of a LangGraph state graph; each reads and appends to a shared typed state (\texttt{ImageState}). The reporting agent emits an annotated overlay, a raw-mask overlay, and a JSON record per plate; aggregated over a batch, these feed the batch-level analysis module, whose \textsc{InsightAgent} uses an LLM and assembles a self-contained HTML report with summary tables, replicate statistics, and a grounded narrative.}
  \label{fig:workflow}
\end{figure*}

The overall workflow is illustrated in Figure~\ref{fig:workflow}. \sys{} was implemented in Python using LangGraph. Each run is initiated with an image path and four metadata fields: \texttt{sample\_id}, \texttt{dilution}, plated \texttt{volume} in mL, and \texttt{replicate\_id}. Each agent reads the shared state, adds its outputs and warnings, and passes the updated state to the next step. A typed dictionary is used to define these records, allowing individual agents to be modified while retaining a consistent interface. The final state provides a serializable record of the analysis. Inputs can be supplied as individual images, folders, or metadata CSV files. Agent roles are defined by their inputs, tools, and outputs; general-purpose agents can follow the same procedure using the analysis tools and shared records.

\subsection{Per-Image Agents}
\paragraph{ImageQCAgent.} Image quality is assessed before segmentation. Focus is evaluated using the variance of the Laplacian, and exposure is assessed by comparing mean brightness with predefined bounds. Plate presence is checked using Hough-circle detection with a contour-based fallback. Failed checks are recorded in \texttt{qc\_status} and reported as warnings; the analysis continues so that users can review questionable images.

\paragraph{PlateSegmentationAgent.} The plate is localized as a circle and the image is masked to retain its interior. This step reduces interference from labels, glare, and the surrounding background. When no plate is detected, the full image is used and the fallback is recorded.

\paragraph{ColonyDetectionAgent.} Colony segmentation is performed using SAM\,3~\cite{carion2025sam3}. Before segmentation, CLAHE is applied to the L channel in Lab color space, followed by mild unsharp masking to improve the visibility of faint colonies. Multiple text prompts are then used, including ``small circular light-brown colonies on TSA agar plate interior'', ``discrete circular colonies on tryptic soy agar'', and ``round isolated colony spots on agar surface''. Masks obtained from these prompts are pooled and filtered using relative area ($2\times10^{-5}$--$1.5\times10^{-2}$ of the image area), circularity ($4\pi A/P^2 \geq 0.2$), and greedy non-maximum suppression (retained IoU $<0.3$). The latter removes duplicate detections from overlapping prompts. Both pooled masks and retained detections are stored for comparison. A SAM~(v1)~\cite{kirillov2023sam} automatic-mask-generator backend is also available for CPU-only use.

\paragraph{ColonyCountingAgent and CFUCalculationAgent.} Colony counts are determined from the number of retained detections. Plates with counts outside the configured 30--300 range are classified as either below counting range or too numerous to count (TNTC). The concentration is calculated as $\mathrm{CFU/mL}=N/(d\cdot V)$, where $N$ is the count, $d$ is the dilution fraction, and $V$ is the plated volume in mL. When required metadata are missing or non-positive, a warning is recorded and no concentration is calculated.

\paragraph{ValidationAgent.} Detection results are further evaluated using rules for overlapping boxes, objects near the plate border, and overgrowth ($\geq 500$ colonies). These checks flag potential merged colonies and edge artifacts. Validation results are stored separately from image quality scores to distinguish image-related concerns from uncertainty in the count.

\paragraph{ReportingAgent.} The reporting agent generates an annotated image showing detections, scores, count, CFU/mL, and quality and validation status. A raw-mask overlay is also produced to show the proposals before filtering. The full state is saved as a per-image JSON record, and the quantitative results are added to a running CSV log.

\subsection{Batch Analysis and Insight Generation}
For folder or CSV runs, results are grouped by \texttt{sample\_id}. The mean, standard deviation, and coefficient of variation are calculated for each group, and pairwise Welch's $t$-tests are performed on counts and CFU/mL. After that, an LLM-based \textbf{InsightAgent} receives the summary table and test results as structured JSON and drafts a results narrative. This role can be assigned to a general-purpose agent that follows the reporting instructions and uses the computed records. The system prompt instructs the model to quote the provided values and limit interpretation to the supplied data. When the model is unavailable, narrative generation is skipped and the remaining report is produced. Summary tables, bar plots with replicate points, test results, the narrative with its model name, and annotated and raw-mask thumbnails are included in a self-contained HTML file for storage and sharing.

\subsection{Skill-Guided Analysis of Study Archives}
The analytical workflow can also be applied to existing microbial population datasets. In an exploratory extension, a general-purpose agent follows a population-analysis skill to inventory files, calculate replicate summaries, reconcile workbook and GraphPad Prism sources, and retain source-cell provenance. The skill distinguishes colony-count reconstruction from workbook arithmetic and re-aggregation of existing CFU or logarithmic values. Detection-limit records remain flagged as bounds; summaries of their stored numbers are identified separately as archive-reproduction statistics. Numerical outputs are calculated by executable scripts before comparison with the published reference values.

%% =====================================================================
\section{Demonstration}
%% =====================================================================
\begin{figure}[t]
  \centering
  \includegraphics[width=0.32\columnwidth]{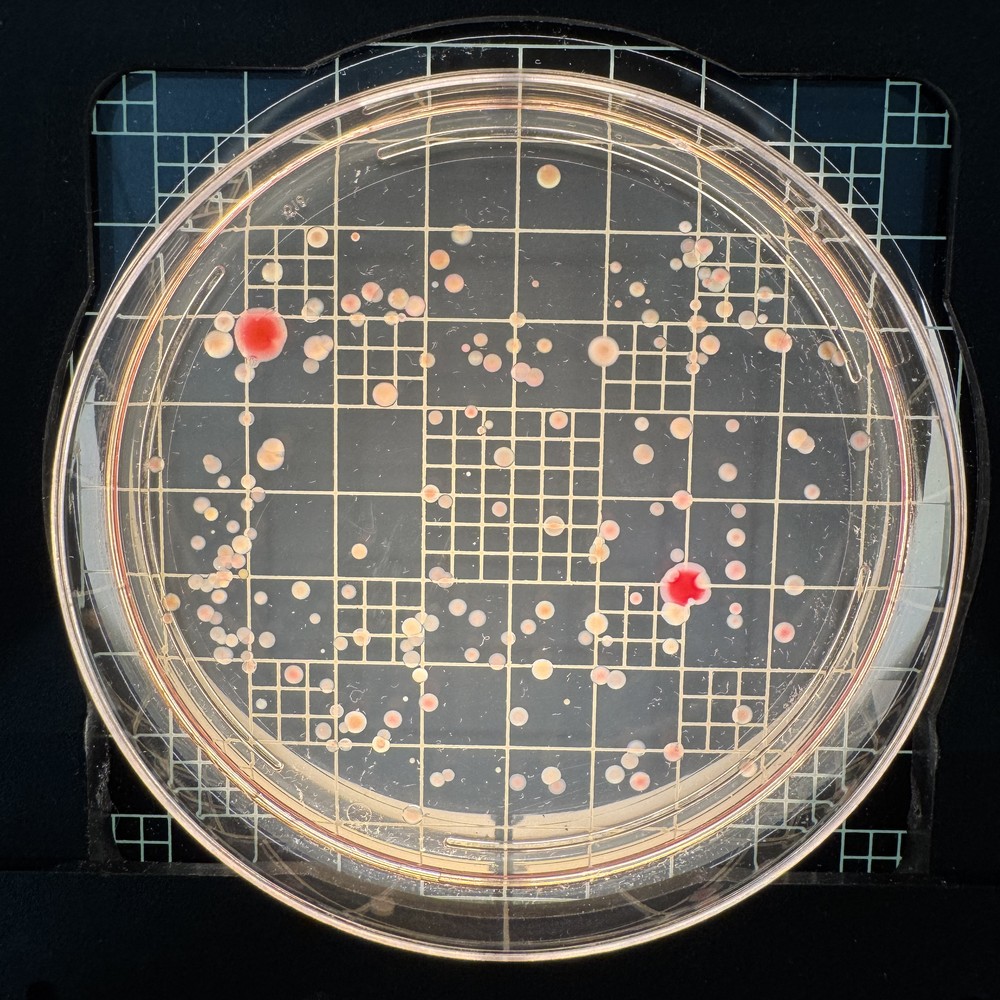}\hfill
  \includegraphics[width=0.32\columnwidth]{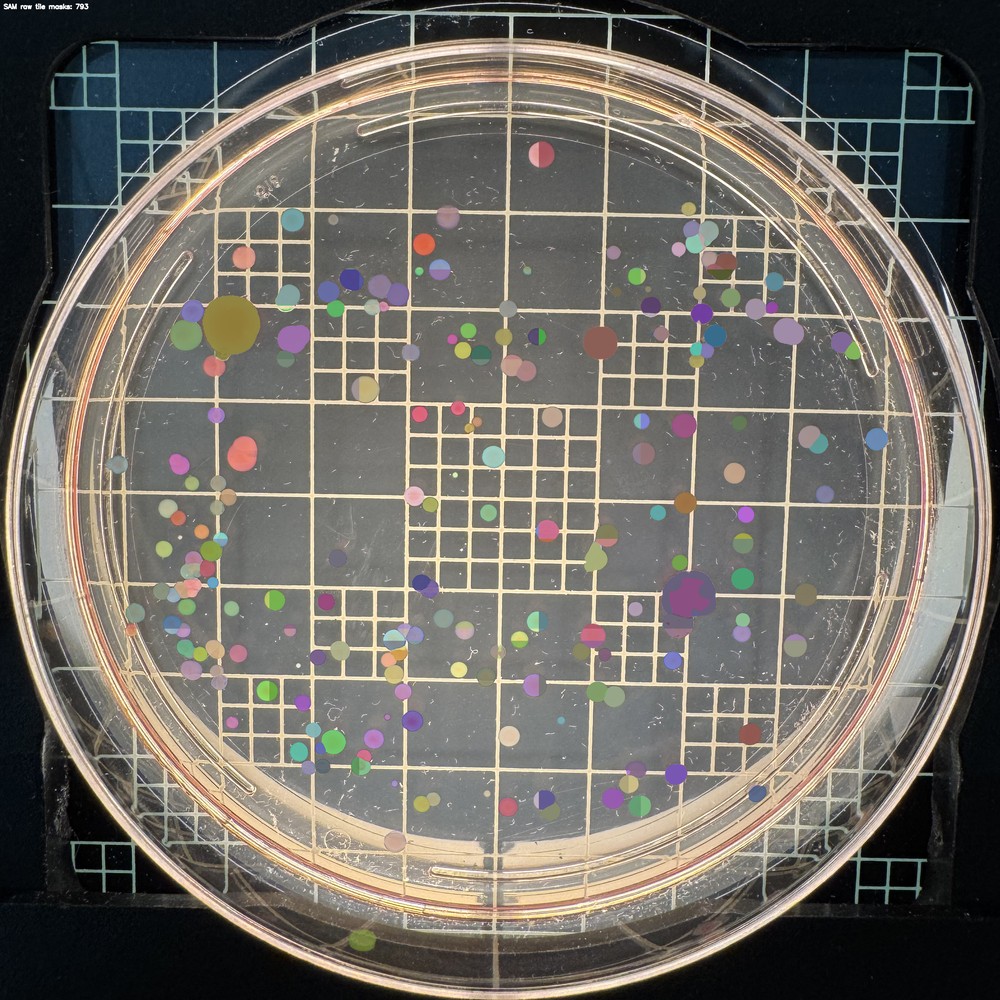}\hfill
  \includegraphics[width=0.32\columnwidth]{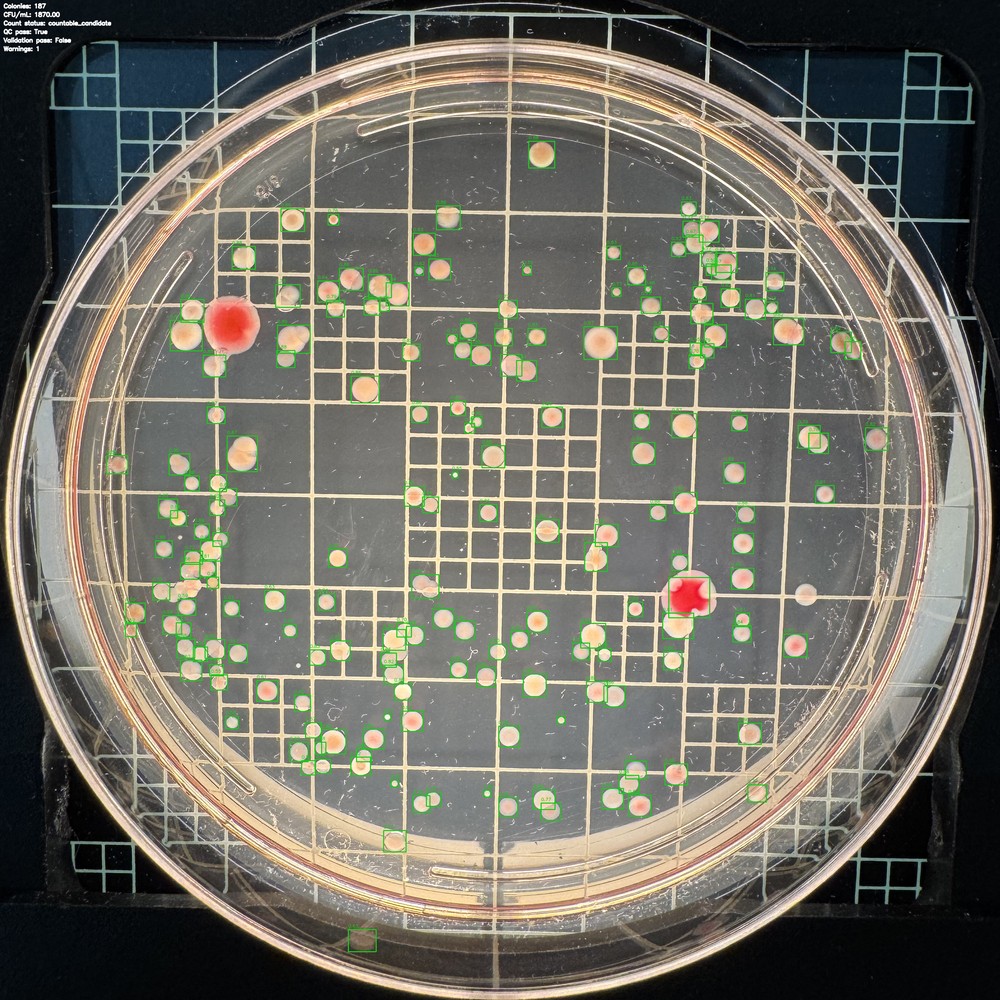}
  \caption{Demonstration of the agent using SAM\,3 to achieve colony counting: input image, raw SAM\,3 masks from the prompt ensemble, and the final annotated output.}
  \label{fig:example}
\end{figure}

\paragraph{Setup.} The proposed demonstration combines the SAM\,3-based image analysis with an LLM-based reporting agent, benchmarked against human-performed counts. We collected 12 MAC, 13 MRS, 21 PCA, and 11 VRBA agar plate images from varied experimental heterogeneous bacterial samples, at multiple plated dilutions. All image processing associated with this analysis were run on a previous generation consumer NVIDIA RTX 3060 GPU with 12GB VRAM. The complete benchmark dataset, comprising 57 high-resolution agar plate images, 2 negative control plates, and corresponding reference annotations, are publicly available through Zenodo (DOI: 10.5281/zenodo.23376472).

\paragraph{Workflow.} Each image was processed through the workflow to generate quality scores, the localized plates, raw segmentation masks, retained detections, and CFU/mL. In detail, images were run with enhancement, tiling, and circularity enabled. Each media type was run with a specific prompt. For example, the MAC agar plates shown in Figure ~\ref{fig:example} were run with the prompt ``Individual round bacterial colonies, cream, pale pink or red, on translucent gray-brown agar. Ignore straight grid lines, scratches, reflections and the dish rim''. This illustrates how language prompts and filtering rules can be used modularly in the pipeline to adapt detection to different plate conditions, without task-specific labeling or retraining.

\paragraph{Results.} Of the fifty-seven images analyzed by \sys{}, 49 (86.0\%) were correctly classified as either countable or outside of the countable range (Table~\ref{tab:results}). For the 29 plates correctly identified by the agent as countable, the accuracy of predicted counts within $\pm 10\%$ of the manual human count, was approximately $89.7\%$. When transformed to a log scale for CFU/ml, this error in predicted count becomes increasingly negligible, as the agent ultimately achieved a CFU/ml accuracy of $ 100 \% $  within $\pm 0.5$~log$_{10}$~CFU/mL of the human derived concentration, across all 29 plates correctly identified as countable. Collectively, these findings demonstrate that \sys{} achieves high concordance with manual colony enumeration for correctly identified countable plates, with overall performance primarily constrained by upstream segmentation and countability classification rather than quantitative estimation of microbial abundance. Further, the segmentation failures were largely restricted to plates with high density of colonies, suggesting that spatial overlap and ambiguous colony boundaries represent key limitations in the model's ability to resolve individual instances under crowded conditions. 

On an NVIDIA RTX 3060 GPU, \sys{} achieved an average end-to-end inference latency of 33.5 seconds per image, corresponding to a processing throughput of approximately 1.8 images per minute and a total execution time of 32 minutes per experimental run. These results demonstrate the feasibility of deploying \sys{} on previous-generation consumer-grade GPU hardware, with modest VRAM requirements and manageable computational overhead.    

\begin{table}[t]
  \centering
  \caption{Countability classification and colony counts versus manual counting on 57 plate images.}
  \label{tab:results}
  \small
  \setlength{\tabcolsep}{4pt}
  \begin{tabularx}{\linewidth}{>{\raggedright\arraybackslash}Xccc}
    \toprule
    \multicolumn{4}{l}{\textbf{Classification}} \\
    \addlinespace[3pt]
    Label
      & \shortstack{Predicted\\Below\\countable}
      & \shortstack{Predicted\\countable}
      & \shortstack{Predicted\\TNTC} \\
    \midrule
    Below countable & 14 & 2  & 0 \\
    Countable       & 2  & 29 & 0 \\
    TNTC            & 0  & 4  & 6 \\
    \midrule
    \multicolumn{4}{l}{\textbf{Count}} \\
    \addlinespace[3pt]
    Mean absolute error
      & \multicolumn{3}{l}{4.3 colonies} \\
    Mean absolute percentage error
      & \multicolumn{3}{l}{4.7\%} \\
    Within $\pm 10\%$ of human count
      & \multicolumn{3}{l}{26/29 (89.7\%)} \\
    \bottomrule
  \end{tabularx}
\end{table}

\paragraph{Archive evaluation against published studies.}
To evaluate analysis beyond image counting, we applied the skill-guided extension to archived datasets from our published dried-peach~\cite{liu2022peaches} and meat-analogue~\cite{liu2023plantmeat} studies. The retrospective, unblinded trial reused the workbook extraction code, recomputed workbook and GraphPad Prism statistics with executable scripts, and compared them against published reference values. Beyond an initial set of 25 selected means (20/25 exact at two decimals; all five peach endpoints exact), we extended the benchmark to a near-complete census of the meat-analogue pathogen tables: all 216 mean$\pm$SD values across three pathogens, three matrices, two agars, and four sampling points at 4, 22, and 32$^\circ$C (Tables~2--4 of~\cite{liu2023plantmeat}), each reconstructed from per-replicate records (Table~\ref{tab:archive}). Every value matched an internal source. On the 168 endpoints with a complete $n{=}6$ source, 131 matched the printed value exactly at two decimals and 164 (97.6\%) agreed within 0.01 log$_{10}$ (MAE 0.0065 log$_{10}$), with only three substantive discrepancies; the 4$^\circ$C table reproduced 67/72 endpoints exactly (MAE 0.0028 log$_{10}$).

The remaining non-matches were informative rather than computational errors. Two 32$^\circ$C \textit{Salmonella} endpoints carried anomalously large published standard deviations (0.46 and 0.81 log$_{10}$) and diverged from the single canonical Prism column, indicating pooled source data; the rest differed only at the last printed digit. Cross-source reconciliation surfaced two completeness findings: a day-0 ground-beef PCA discrepancy between workbook arithmetic (5.6088) and Prism logs (5.4426), and a \textit{Listeria} abuse-temperature series (48 endpoints) whose archived file holds only one of the two trials the paper pooled ($n{=}3$ vs printed $n{=}6$), reported separately rather than as errors. Compared endpoints were re-aggregations of existing values, rather than independent colony-count reconstructions. This evaluates the archive-analysis extension separately from image-counting accuracy.

\begin{table}[t]
  \caption{Agreement of recomputed archive values with 241 published endpoints.}
  \label{tab:archive}
  \small
  \centering
  \setlength{\tabcolsep}{4pt}
  \begin{tabular}{lccc}
    \toprule
    Source & Endpoints & Exact 2\,d.p. & $|\Delta|{\leq}0.05$ \\
    \midrule
    Selected microflora + peach & 25 & 20 & 23 \\
    Pathogens, 4$^\circ$C & 72 & 67 & 72 \\
    Pathogens, 22$^\circ$C (\textit{Ec}, \textit{Sm}) & 48 & 36 & 47 \\
    Pathogens, 32$^\circ$C (\textit{Ec}, \textit{Sm}) & 48 & 28 & 46 \\
    \textit{Listeria}, 22/32$^\circ$C ($n{=}3$) & 48 & 5 & 25 \\
    \midrule
    Combined & 241 & 156 & 213 \\
    \bottomrule
  \end{tabular}
\end{table}

%% =====================================================================
\section{Discussion and Outlook}
%% =====================================================================
This study shows that an agentic workflow can fulfill the routine colony-counting work of a microbiologist while producing a more reliable, auditable record. \sys{} placed every correctly identified countable plate within $\pm 0.5$ log$_{10}$ CFU/mL of the manual result (Table~\ref{tab:results}), and every reported value is traceable to its masks, detections, and warnings, so a reviewer can see which objects were counted and why a plate was accepted or rejected. In the archive evaluation, recomputed statistics matched 97.6\% of complete published endpoints and surfaced source-data inconsistencies missed in manual analysis.

Manual enumeration has been reported to take approximately 4 minutes per plate~\cite{frost2016counting}; at that rate, the fifty-seven plates in our benchmark would require about 3.8 hours of hands-on counting, whereas \sys{} processed them unattended in 32 minutes (33.5 seconds per image), roughly a 7-fold reduction even before CFU calculation, replicate statistics, and reporting, which \sys{} also automates and whose savings grow with the number of conditions, time points, dilutions, and replicates. Commercial automated colony counters also reduce counting time, but they typically cost several thousand US dollars and rely on fixed, hard-coded detection algorithms; in contrast, \sys{} is open source, runs on a previous-generation consumer GPU (NVIDIA RTX 3060, 12\,GB VRAM) with ordinary plate photographs, and adapts to new media through text prompts rather than retraining or vendor-specific configuration. Remaining errors were concentrated on crowded plates with merged or edge colonies (Figure~\ref{fig:example}), motivating targeted re-segmentation with point prompts, and a verifier agent that checks LLM narratives against computed statistics is a further direction.

\bibliographystyle{ACM-Reference-Format}
\bibliography{refs}

\end{document}